\def\optLevelONE{}
\documentclass[conference]{IEEEtran}
\usepackage{cite}
\usepackage{atbegshi}
\usepackage{inputenc}
\renewcommand{\baselinestretch}{0.98}
\usepackage[pdftex,dvipsnames,usenames,dvipsnames,svgnames,table]{xcolor}  
\definecolor{color0}{RGB}{255,0,0}
\definecolor{color1}{RGB}{0,255,0}
\definecolor{color2}{RGB}{102,140,217}
\definecolor{color3}{RGB}{16,150,24}
\definecolor{color4}{RGB}{153,0,153}
\definecolor{color5}{RGB}{255,165,0}
\definecolor{cieee0}{HTML}{00629b}
\definecolor{cieee1}{RGB}{255,199,44}
\definecolor{cieee2}{RGB}{232,119,34}
\definecolor{cieee3}{RGB}{186,12,47}
\definecolor{cieee4}{RGB}{119,37,131}
\definecolor{cieee5}{RGB}{120,190,32}
\definecolor{cieee6}{RGB}{0,132,61}
\definecolor{cieee7}{RGB}{0,159,223}
\definecolor{BLACK}{RGB}{255,255,255}
\usepackage{changes}
\renewcommand{\deleted}[1]{}
\newcommand{\addedfinal}[1]{#1}
\newcommand{\deletedfinal}[1]{}

\newcommand{\shortOrLong}[2]{#1}
\usepackage{scrlayer-scrpage}
\usepackage{multicol}
\usepackage{multirow}
\usepackage{lipsum}
\usepackage{fp}

\usepackage{array}
\usepackage{siunitx}

\newcommand{\cerutti}{Cerutti et al.}
\newcommand{\calc}[2]{\FPeval{\calcresult}{round(#2,#1)}\calcresult}

\usepackage{xargs}  
\newcommandx{\info}[2][1=]{\todo[linecolor=green,backgroundcolor=green!25,bordercolor=green,#1]{#2}}
\newcommandx{\unsure}[2][1=]{\todo[linecolor=red,backgroundcolor=red!25,bordercolor=red,#1]{#2}}
\newcommandx{\change}[2][1=]{\todo[linecolor=blue,backgroundcolor=blue!25,bordercolor=blue,#1]{#2}}
\newcommandx{\improvement}[2][1=]{\todo[linecolor=Plum,backgroundcolor=Plum!25,bordercolor=Plum,#1]{#2}}
\newcommand{\hide}[1]{}
\usepackage{amsmath}
\usepackage{blindtext, graphicx}
\usepackage{multirow}
\usepackage{tikz,pgfplots,pgfplotstable}
\usepackage{ulem}
\usepackage[bookmarks=false,hidelinks]{hyperref}
\usepackage{textgreek}
\pgfplotsset{compat=1.5,width=10cm}
\usepgfplotslibrary{external}
\usepackage{verbatim}
\colorlet{A1}{red!80!white}
\colorlet{B1}{orange!80!white}
\colorlet{C1}{blue!80!white}
\colorlet{A2}{red!80!black}
\colorlet{B2}{orange!80!black}
\colorlet{C2}{blue!80!black}

\newcommand{\iscasrev}[1]{#1}

\usepackage{lipsum}
\usepackage{ifthen}
\usepackage[font=small]{caption}
\usepackage{pdftexcmds}
\usepackage{stringstrings}
\usepackage{multicol}
\usepackage{graphicx}
\usepackage{booktabs}
\usepackage{pgfplots}
\pgfplotsset{compat=1.13}
\usepackage{pgfplotstable}

\usepackage{algorithm}
\usepackage[noend]{algorithmic}
\usepackage{tabularx}
\usepackage{amsfonts}
\let\labelindent\relax
\usepackage{enumitem}
\usepackage{xspace}

\definecolor{bblue}{HTML}{4F81BD}
\definecolor{rred}{HTML}{C0504D}
\definecolor{ggreen}{HTML}{9BBB59}
\definecolor{ppurple}{HTML}{9F4C7C}
\newcommand{\x}{\ensuremath{\times}\xspace}

\newcommand{\figref}[1]{Fig.~\ref{#1}}

\newcommand{\xnorbinfigures}{./figures/}

\newcommand{\Tsukiji}{ChewBaccaNN{}}
\newcommand{\xnorbin}{\Tsukiji\xspace}
\newcommand{\newiscas}[1]{#1}

\newcommand{\rmifneeded}[1]{}
\hypersetup{draft}

\begin{document}

\bstctlcite{IEEEexample:BSTcontrol}

\title{\fontsize{22pt}{22pt}\selectfont ChewBaccaNN: A Flexible 223 TOPS/W BNN Accelerator}

\ifdefined\noack
\else
\author{\IEEEauthorblockN{Renzo Andri\IEEEauthorrefmark{4}\IEEEauthorrefmark{1}, Geethan~Karunaratne\IEEEauthorrefmark{1}\IEEEauthorrefmark{2}, Lukas Cavigelli\IEEEauthorrefmark{4}\IEEEauthorrefmark{1}, Luca Benini\IEEEauthorrefmark{1}\IEEEauthorrefmark{3}}\\
\IEEEauthorblockA{\IEEEauthorrefmark{1}Integrated Systems Laboratory, ETH Zurich, Zurich, Switzerland \IEEEauthorrefmark{2}IBM Research, Zurich, Switzerland \\\IEEEauthorrefmark{4}Huawei Technologies, Zurich Research Center, Zurich, Switzerland
\IEEEauthorrefmark{3}DEI, University of Bologna, Bologna, Italy}}
\fi
\IEEEoverridecommandlockouts
\IEEEpubid{\parbox[t]{\columnwidth}{\vspace{-23mm}\textcopyright 2021 IEEE.  Personal use of this material is permitted. Permission from IEEE must be obtained for all other uses, in any current or future media, including reprinting/republishing this material for advertising or promotional purposes, creating new collective works, for resale or redistribution to servers or lists, or reuse of any copyrighted component of this work in other works. \newline
IEEE ISCAS 2021, Daegu, South Korea, 23--26 May 2021\\
DOI: TBD\\URL: http://ieeexplore.ieee.org/document/TBD/\hfill} \hspace{\columnsep}\makebox[\columnwidth]{ }}
\maketitle 
\IEEEpubidadjcol
\begin{abstract}
\newiscas{Binary Neural Networks enable smart IoT devices, as they significantly reduce the required memory footprint and computational complexity while retaining a high network performance and flexibility. This paper presents \xnorbin, a 0.7\,mm\textsuperscript{2} sized binary convolutional neural network (CNN) accelerator designed in GlobalFoundries 22\,nm technology. By exploiting efficient data re-use, data buffering, latch-based memories, and voltage scaling, a throughput of \calc{0}{241.1565} GOPS is achieved while consuming just 1.1\,mW at 0.4V/154MHz during inference of binary CNNs with up to 7$\times$7 kernels, leading to a peak core energy efficiency of \calc{0}{222.93333}\,TOPS/W. \xnorbin's flexibility allows to run a much wider range of binary CNNs than other accelerators, drastically improving the accuracy-energy trade-off beyond what can be captured by the TOPS/W metric. In fact, it can perform CIFAR-10 inference at 86.8\% accuracy with merely 1.3\,\textmu J, thus exceeding the accuracy while at the same time lowering the energy cost by 2.8$\times$ compared to even the most efficient and much larger analog processing-in-memory devices, while keeping the flexibility of running larger CNNs for higher accuracy when needed. It also runs a binary ResNet-18 trained on the 1000-class ILSVRC dataset and improves the energy efficiency by \calc{1}{222.93333/50.6}$\times$ over accelerators of similar flexibility. Furthermore, it can perform inference on a binarized ResNet-18 trained with 8-bases Group-Net to achieve a 67.5\% Top-1 accuracy with only 3.0\,mJ/frame---at an accuracy drop of merely 1.8\% from the full-precision ResNet-18. }
\end{abstract}


\IEEEpeerreviewmaketitle

\section{Introduction}
\IEEEPARstart{C}{NNs} have revolutionized the ML field in recent years, outperforming humans in image recognition \cite{He2015} and advancing the SoA for a wide range of applications \cite{khan2019survey}. Most of these networks require billions of multiply-accumulate (MAC) operations per frame and millions of trained parameters. This is incompatible with the few hundreds of kB of on-chip memory and the limited energy available on battery-powered, low-cost IoT sensor nodes.

Sending the data to the cloud for analysis seems like a viable option to forego these challenges. However, transmitting the data comes at a high energy cost, introduces privacy concerns, requires expensive infrastructure, and has high latency. Alternatively, the challenge of analyzing the data near the sensor can be tackled by a combination of algorithmic optimizations to allow working with reduced arithmetic precision, and hardware acceleration \cite{Sze2017,Andri2016a} with various techniques \newpage\noindent maximize energy efficiency, such as minimizing off-accelerator data transfers \cite{AndriHyperdriveJETCAS, moini2017resource, 8886603}.

Reducing the arithmetic precision has shown significant potential, affecting both the complexity of the compute operation itself as well as reducing data movement and storage requirements for the feature maps and the weights. 

Inference with 8\,bit operands has become common-place due to the negligible accuracy loss for many applications with support available in many deep learning frameworks and commercial hardware \cite{jacob2018quantization}. 

At the extreme limit, binarized neural networks (BNNs) quantize weights and feature maps to a single bit representing the values -1 and 1. XNOR-Net extends the stochastic gradient descent algorithm (commonly used to train NNs) by quantizing the weights and activations in the forward path and scales the feature maps $\ell^1$ matrix norm of the weight kernels \cite{Rastegari2016}. On the challenging ImageNet dataset, Rastegari et al. achieved 51.2\% using a binarized ResNet-18, which was a significant drop of -18.1\%. Courbariaux et al. achieved then SoA results with 99.04\% on MNIST (+0.34\%), 97.47 on SVHN (-0.09\%), and 89.85\% CIFAR-10 (-0.46\%), but these datasets are much simpler than ImageNet \cite{hubara2016binarized}.

Recent research has been focusing mainly on minimizing the quantization error (e.g., scaling feature maps in XNOR-Net), improving the loss function, and reducing the gradient error \cite{qin2020binary, spallanzani2019additive}. Recently, the accuracy gap between BNNs and their full-precision equivalents have been brought down to 12\% (DoReFa-Net on Alexnet \cite{zhou2016dorefa}) and can be reduced by using multiple binary layers (i.e., weight bases) in parallel and binarizing around multiple thresholds (i.e., activation bases). Using 3 weight bases, Lin et al. \cite{lin2017towards} have achieved 69.3\%/89.2\% (Top-1/Top-5, --8.3\%/--6.0\% vs. ResNet-18) and Zhuang et al. reached 72.8\%/90.5 (Top-1/Top-5, --3.2\%/--2.4\% vs. ResNet-50) using 8 weight bases \cite{zhuang2019structured}. Using multiple bases directly impacts the throughput and energy per inference for any hardware accelerator, negating some of the benefits of BNNs. However, it enables smoothly scaling from a highly efficient, less accurate network to almost full accuracy inference. 

\newiscas{BNN hardware accelerators have demonstrated around two orders of magnitude energy efficiency gain compared to quantized NN accelerators by avoiding off-chip data transfers and exploiting extremely reduced arithmetics where a MAC becomes binary \textit{xnor}-popcount. Conti et al. present a 46\,TOPS/W BNN accelerator tightly-connected to a general-purpose processor (omitting I/O costs) \cite{conti2018xnor}, UNPU is a stand-alone accelerator for flexible weights (i.e., 1-16\,bit) and feature maps and reaches 51\,TOPS/W for fully-binary NN \cite{lee2018unpu}. Furthermore, accelerators with fixed and uncommon kernel sizes $2\times 2$ \cite{bankman2018alwaysCORRECT, knag2020617} and analog computation \cite{bankman2018alwaysCORRECT, valavi201964} show a theoretic peak energy efficiency of 866\,TOPS/W, but lack flexibility and require significant larger BNNs for competitive network performance.}

In this paper, we present \xnorbin, an enhanced architecture and novel implementation \iscasrev{based on} XNORBIN \cite{Bahou2018N}, increasing the energy efficiency by \calc{1}{223/95}$\times$ through \iscasrev{a redesign} of the memory architecture including a switch to latch-based memories to allow more aggressive voltage scaling down to 0.4V and adjusted memory sizes to eliminate off-chip storage of intermediate results, as well as enhanced power gating to dynamically turn off unused memory banks, and a more advanced technology node. Combining these methods, we achieve an energy efficiency of 223\,TOPS/W with a fully-digital accelerator in 22\,nm. New features, including support for average pooling and residual paths, dramatically boost the accuracy-energy trade-off beyond the scope of the TOPS/W metric, allowing us to run inference at 91.5\% accuracy with \SI{7.3}{\micro\joule} or 86.8\% with \SI{1.3}{\micro\joule} on CIFAR-10---an improvement of 2.8$\times$ over the state-of-the-art.

\section{BNN and Related HW Optimization}
\label{sec:bin_conv}
In BNNs, the weights and intermediate feature maps are quantized to a single bit:  $\textbf{I} \in \{-1, 1\}^{n_{in}\times i_h\times i_w}$ $\textbf{W} \in \{-1, 1\}^{n_{out}\times n_{in}\times h_k \times w_k} $, with spatial feature map size of $i_h\times i_w$, kernel size $h_k\times w_k$ and number of input/output channels $n_{in}$, $n_{out}$. After the multiplication, which is reduced to an \texttt{xnor} operation, these products and a bias value are accumulated in full-precision followed by re-binarizing the sum. This (re-)binarization replaces the activation function and has the behavior of a signum function while zero is mapped to 1, i.e., $\forall x>0 : \text{sgn}(x)=1$ and $\forall x\le0 : \text{sgn}(x)=-1$. The $\ell$-th output feature map $\mathbf{o_\ell}$ is the sum of convolutions of every binarized input feature map $\mathbf{{i}_n}$ with the corresponding binary weights $\mathbf{{w}_{\ell,n}}$ and the bias $C_\ell$:
\begin{align}
\mathbf{o_\ell} &= \text{sgn}\left(C_\ell+\alpha_\ell\sum_{n=0}^{n_{in}-1}{{\text{sgn}(\mathbf{i_n}) \ast \text{sgn}(\mathbf{w_{\ell,n}})}}\right)\label{eq:bnn_noopt}
\end{align}
BNNs have much potential for optimizations: First, the memory footprint can be reduced up to 32\x{} (relative to FP32). Second, multiplications can be simplified to \texttt{xnor} operations. Training of BNNs is not trivial, as signum function is not smooth and thus not differentiable. The most common approach, known as straight-through estimator, is based on shadow weights in high precision (e.g., FP32) \cite{bengio2013straightthroughestimator}. These weights are binarized during the forward-propagation while during back-propagation, the gradients are passed on to the input as without quantization and the gradient update is applied to the shadow weights. Typically, the weights of the first and last layers are not binarized, as this has a strong impact on the network performance, but contributes a small part of the total compute effort \cite{zhou2016dorefa}.

For implementation, the bipolar activations and weights $i_n, w_{\ell,n}\in\{-1, 1\}$ are mapped to the binary representation $\hat{i}_n, \hat{w}_{\ell,n}\in\{0, 1\}$ with $\hat{x}=$ $\frac{1}{2}(x$\,$+$\,$1)$, enabling the replacement of the multiplication with a \texttt{xnor} operation. This introduces an offset ($-h_k w_k$) and scaling factor ($2$) that need to be applied before the re-binarization:

\resizebox{0.97\linewidth}{!}{
  \begin{minipage}{\linewidth}

  \begin{align*}
\mathbf{o_\ell} &= \text{sgn}\left(C_\ell+ \alpha_\ell\sum_{n=0}^{{n_{in}}-1}{\left(2\cdot\mathbf{\hat{i}_{n}} \ast^{\bar{\oplus}} \mathbf{\hat{w}_{\ell,n}}-h_k w_k\right)}\right) 
\end{align*}
  \end{minipage}
}

\noindent The factor and offset of the BNN convolution, $C_\ell$, $\alpha_\ell$, and the parameters $\mu_\ell$, $\sigma_\ell$ of the following batch normalization layer can be absorbed into single threshold $\theta_\ell=\frac{\hat{C}_\ell\sigma_\ell}{|\hat{\alpha}_\ell|}+\mu_\ell$, applied to the sum of products,

\begin{align}
\mathbf{o_\ell} = 
\begin{cases}
-1, & \sum_{n=0}^{{n_{in}}-1}{\mathbf{\hat{i}_{n}} \ast^{\bar{\oplus}} \mathbf{\hat{w}_{\ell,n}}}<\theta_\ell\\
1, & \text{else}
\end{cases}.
\end{align}

\noindent Pooling is applied after convolution, scaling, and batch normalization, but before the re-\hspace{-0.25mm}binarization, therefore, in the non-binary domain. However, due to the monotonicity and commutativity, the pooling can be calculated as a Boolean operation (e.g., max/min-pooling becomes \texttt{AND}/\texttt{OR} reduction).

\ifdefined\optLevelONE\else
\begin{align}
Pool(o_\ell(x,y)) &= 
\begin{cases}
-1, & \max\limits_{m,n\in\{0,1\}}(o_\ell(2x+m,2y+n))<\theta_\ell\\
1, & else
\end{cases}\\
 &= 
\begin{cases}
-1, & \bigwedge\limits_{m,n\in\{0,1\}}\left(o_\ell(2x+m,2y+n)<\theta_\ell\right)\\
1, & else
\end{cases}
\end{align}\fi

\section{Architecture}
\begin{figure*}
        \centering \includegraphics[width=\linewidth]{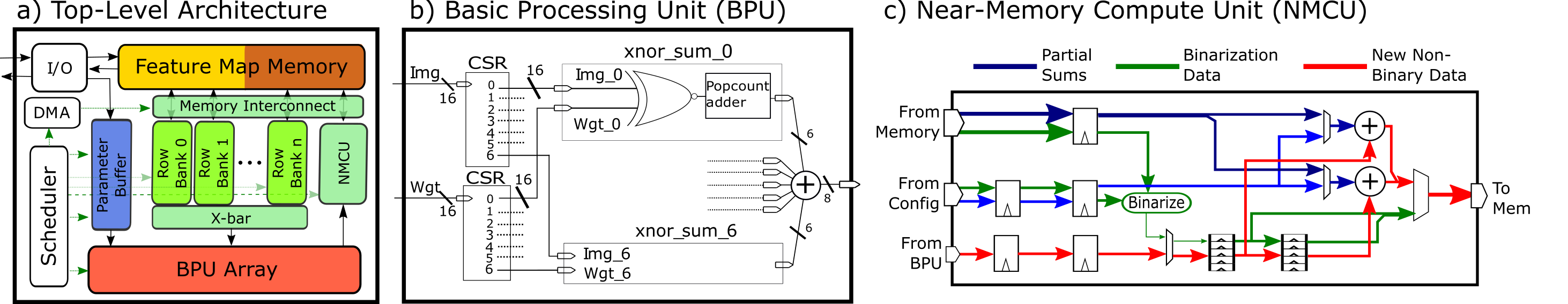}
        \captionof{figure}{Architecture overview. a) Top-level schematic, b) Detailed schem. of a single BPU, c) Detailed schem. of near-memory compute unit.}

          \label{fig:architecture_merge}
\end{figure*}
The architecture of \xnorbin{} is illustrated in \figref{fig:architecture_merge}a and its components are explained as follows:

Each \textit{Basic Processing Unit (BPU)} performs a 1D convolution of an input image row with a kernel row from 16 input channels at a time, by employing \textit{xnor\_sum} units consisting of 16 \textit{xnor}-gates each and a popcount adder tree.
The \textit{xnor\_sum} unit is replicated 7 times to produce outputs corresponding to a window size of at most 7 input feature map pixels in a row. Outputs from all units are accumulated together with a second stage adder tree to create a 1D inner product, shown in Fig.~\ref{fig:architecture_merge}b). 7 BPUs \iscasrev{form the BPU array, }instantiated in order to support kernel sizes up to 7$\times $7. The outputs of all these instances are pipelined to increase throughput and are then added up in a third stage adder tree to produce 2D inner product. Each of the \textit{xnor}-sum instances is fed with the input activations and weight data through a controlled shift-register to enable data reuse. The same BPU array datapath is reused to perform binary max-pooling operation, where the 2\textsuperscript{nd} and 3\textsuperscript{rd} stage adder trees are flanked by a 1\,bit comparator (AND gate) tree.
\xnorbin{} comes with a \textit{Feature Map Memory FMM} and data buffering.

\noindent The FMM stores the feature maps and the partial sums of the convolutions. The memory is divided into two blocks, where one serves as the data source (i.e., current input feature maps), and the other serves as data sink (i.e., partial or final output feature maps). They are swapped after each layer. If the FMM is dimensioned to fit the largest intermediate FMs, no energy-costly off-chip memory accesses are needed during inference. To hide the weight loading latency, the PB is enriched with a double buffering feature. The \textit{Parameter Buffer PB} stores the weights, the binarization thresholds, and the configuration parameters. In the optimal case, it stores all the weights of the network to avoid I/O for weight loading. If the parameters are too many to fit on-chip, the PB can be reused to buffer off-chip accesses.
    The \textit{Row Banks} are used to buffer rows of the input feature maps for frequent accesses. It also contains rows of filter weights corresponding to the batch of output channels calculated in parallel. Since these row banks need to be rotated when shifting the convolution window down, they are connected to the BPU array through a crossbar. 
    
    The crossbar connects the registers inside the BPUs, the \textit{controlled shift registers} (L1) (CSRs) containing kernel input feature map elements, and the filter weight elements. These are shifted when the convolution window is moved forward. 
     The \textit{DMA} moves data independently from FMM and PB into (via Row Banks) and out of the BPU array. 
     \textit{Scheduler}: According to the layer configuration of a CNN, the scheduler controls the crossbar on how and when to route feature map and weight data from the Row Banks to the BPUs in order to compute row-wise partial sums for each member in the batch.
     The \textit{Near Memory Compute Unit (NMCU)} is shown in Fig.~\figref{fig:architecture_merge}c and is used for on-the-fly computation when writing back to the FFM. This includes partial sum calculations from the BPU array\addedfinal{, accumulating residual paths from the FMM}, the binarization, and storing in FMM in a packed format (i.e., 16 activations). 

To maximize kernel-level reuse, filter weights are retained in BPUs while streaming selected image rows through BPUs, and partial sums are computed concurrently for a batch of output feature map tile to maximize row-level image reuse.
The resulting integer value that is produced from the BPU array in each cycle as a result of a horizontally sliding convolution window is forwarded to the DMA controller via the \textit{Near-Memory Compute Unit (NMCU)}. The NMCU accumulates the partial results by means of a read-add-write operation. After the final accumulation of partial results, the unit also performs the thresholding/re-binarization operation (i.e., activation and batch normalization). The binary results are packed into 16\,bit words and written back to the memory by the DMA unit.
The scheduling is determined with the objective of maximizing the data reuse at different levels of the memory hierarchy.

The scheduling algorithm and mapping of operations to BPU units are explained in Alg.~\ref{algo:conv_schedule} based on the filter dimensions $k_w$ and $k_h$, the spatial input dimensions $i_w$ and $i_h$, the depths (i.e., input channels $c_i$, and output channels $c_o$) and the channels tile sizes $\hat{c}_i$ and $\hat{c}_o$. Parallel execution is indicated in the brackets in lines 4, 6---8. After one tile of output channels $\hat{c}_o$ is computed (i.e., binary convolution, (optional) pooling and thresholding), the process is repeated for all output channel tiles. In the next step, the next subsequent layer with new layer configuration is loaded, the FMM sink/source direction is reversed, and calculated.

\begin{algorithm}
  \caption{High-level scheduling of 1 BNN layer}\label{algo:conv_schedule}
\begin{algorithmic}[1]
\ifdefined\arxiv\else\footnotesize\fi

\REQUIRE{ $k_w, k_h, i_w, i_h, c_i, c_o, \hat{c}_i, \hat{c}_o$}
    \FOR{$n_o\leftarrow 0$ \TO \(c_o\)/$\hat{c}_o $}
        \FOR{$n_i\leftarrow 0$ \TO \(c_i\)/$\hat{c}_i $}
            \STATE pass $n_i^{th}$ chunk of next $\hat{c}_o $ filters to Row Banks
            \FOR{$n_{r}\leftarrow 0$ \TO $i_h $}
                \STATE pass Feature Map (:,$n_{r}$,$n_i\text{*}b_i\text{:+}b_i$) to Row Banks
                \FOR{$b_o\leftarrow 0$ \TO $\hat{c}_o $ }
                    \STATE pass $n_i^{th}$ chunk of $b_o^{th} $ filter to BPU array
                    \FOR{$n_{c}\leftarrow 0$ \TO $i_w $}
                        \STATE pass Feature Map ($n_{c}$,${n_{r}\text{+}k_{r}}$,$n_i\text{*}b_i\text{:+}b_i$) to BPU array
                        \FOR{$k_{r}\leftarrow \text{-}(k_h/2) $ \TO $(k_h/2) $ (parallel in BPU array)}
                            \FOR{$k_{c}\leftarrow \text{-}(k_w/2) $ \TO $(k_w/2) $ (parallel in BPU)}
                                \FOR{$b_i\leftarrow 0$  \TO $\hat{c}_i $ (parallel in BPU gates)}
                                    \STATE Produce Partial Sum ($n_{c}$,$n_{r}$,$\hat{c}_o\text{*}n_o\text{+}b_o$)
                                \ENDFOR
                            \ENDFOR
                        \ENDFOR
                        \STATE Accumulate Partial Sum ($n_{c}$,$n_{r}$,$\hat{c}_o\text{*}n_o\text{+}b_o$) at NMCU
                    \ENDFOR
                \ENDFOR
            \ENDFOR
        \ENDFOR
        \STATE Binarize final Partial Sum (:,:,$n_i\text{*}\hat{c}_i\text{:+}\hat{c}_i$)
        \STATE Pool operation (if applicable)
    \ENDFOR
\end{algorithmic}

\end{algorithm}

\def\plotsymbSEVEN{*}
\def\plotsymbFIVE{triangle*}
\def\plotsymbTHREE{star}

\begin{figure}
\centering
\begin{tikzpicture}[scale=0.75]
\tikzset{mark options={solid, mark size=3, line width=1pt}} %



	\begin{axis}[width=1.33\columnwidth-0.5cm, height=\shortOrLong{6.0cm}{8cm}, 
		xlabel={Throughput {[GOPS]}},ylabel={Core Energy Effic. {[TOPS/W]}},  grid=major, xmin=0, xmax=350, ymin=0, ymax=270]

\addplot[color=cieee1,mark=\plotsymbSEVEN,only marks] table [x=throughput, y expr=\thisrow{eneff}, col sep=space, unbounded coords=discard, restrict expr to domain={\thisrow{technology}}{770015:770915}]{./plots/throughputtable.tex}; \label{soaPlot:7t5.15}
\addplot[color=cieee1,mark=\plotsymbSEVEN,only marks] table [x=throughput, y expr=\thisrow{eneff}, col sep=space, unbounded coords=discard, restrict expr to domain={\thisrow{technology}}{770020:770920}]{./plots/throughputtable.tex}; \label{soaPlot:7t5.20}
\addplot[color=cieee1,mark=\plotsymbSEVEN,only marks] table [x=throughput, y expr=\thisrow{eneff}, col sep=space, unbounded coords=discard, restrict expr to domain={\thisrow{technology}}{770040:770940}]{./plots/throughputtable.tex}; \label{soaPlot:7t5.40}
\addplot[color=cieee1,mark=\plotsymbSEVEN,only marks] table [x=throughput, y expr=\thisrow{eneff}, col sep=space, unbounded coords=discard, restrict expr to domain={\thisrow{technology}}{770080:770980}]{./plots/throughputtable.tex}; \label{soaPlot:7t5.80}

\addplot[color=cieee2,mark=\plotsymbFIVE,only marks] table [x=throughput, y expr=\thisrow{eneff}, col sep=space, unbounded coords=discard, restrict expr to domain={\thisrow{technology}}{550015:550915}]{./plots/throughputtable.tex}; \label{soaPlot:5t5.15}
\addplot[color=cieee2,mark=\plotsymbFIVE,only marks] table [x=throughput, y expr=\thisrow{eneff}, col sep=space, unbounded coords=discard, restrict expr to domain={\thisrow{technology}}{550020:550920}]{./plots/throughputtable.tex}; \label{soaPlot:5t5.20}
\addplot[color=cieee2,mark=\plotsymbFIVE,only marks] table [x=throughput, y expr=\thisrow{eneff}, col sep=space, unbounded coords=discard, restrict expr to domain={\thisrow{technology}}{550040:550940}]{./plots/throughputtable.tex}; \label{soaPlot:5t5.40}
\addplot[color=cieee2,mark=\plotsymbFIVE,only marks] table [x=throughput, y expr=\thisrow{eneff}, col sep=space, unbounded coords=discard, restrict expr to domain={\thisrow{technology}}{550080:550980}]{./plots/throughputtable.tex}; \label{soaPlot:5t5.80}

\addplot[color=cieee3,mark=\plotsymbTHREE,only marks] table [x=throughput, y expr=\thisrow{eneff}, col sep=space, unbounded coords=discard, restrict expr to domain={\thisrow{technology}}{330015:330915}]{./plots/throughputtable.tex}; \label{soaPlot:3t5.15}
\addplot[color=cieee3,mark=\plotsymbTHREE,only marks] table [x=throughput, y expr=\thisrow{eneff}, col sep=space, unbounded coords=discard, restrict expr to domain={\thisrow{technology}}{330020:330920}]{./plots/throughputtable.tex}; \label{soaPlot:3t5.20}
\addplot[color=cieee3,mark=\plotsymbTHREE,only marks] table [x=throughput, y expr=\thisrow{eneff}, col sep=space, unbounded coords=discard, restrict expr to domain={\thisrow{technology}}{330040:330940}]{./plots/throughputtable.tex}; \label{soaPlot:3t5.40}
\addplot[color=cieee3,mark=\plotsymbTHREE,only marks] table [x=throughput, y expr=\thisrow{eneff}, col sep=space, unbounded coords=discard, restrict expr to domain={\thisrow{technology}}{330080:330980}]{./plots/throughputtable.tex}; \label{soaPlot:3t5.80}

\draw[dashed] (305.237,0) -- (305.237,182.231) -- (285.091,182.231) -- (285.091,211.993) -- (261.333,211.993) -- (261.333,213.254) -- (241.231,213.254) -- (241.231,223.149) -- (0,223.149);
\draw[dashed] (155.733,0) -- (155.733,100.427) -- (145.455,100.427) -- (145.455,117.619) -- (133.333,117.619) -- (133.333,119.225) -- (123.077,119.225) -- (123.077,124.365) -- (0,124.365);
\draw[dashed] (56.064,0) -- (56.064,52.755) -- (52.364,52.755) -- (52.364,61.291) -- (48.000,62.339) -- (44.308,65.273) -- (0,65.273);



    	\end{axis}

\end{tikzpicture}\\
\begin{scriptsize}	
$k_x\times k_y=7\times 7$  (\ref{soaPlot:7t5.15}), $5\times 5$ (\ref{soaPlot:5t5.15}), $3\times 3$ (\ref{soaPlot:3t5.15}) 
		 \end{scriptsize}

\caption{{Throughput vs. Core Energy Efficiency for various timing constraints at 0.4\,V supply voltage and FMM=4\,kB.}\ifdefined\arxiv\else\vspace{-7mm}\fi}\label{fig:xnorbinenergyvsthroughput}
\end{figure}
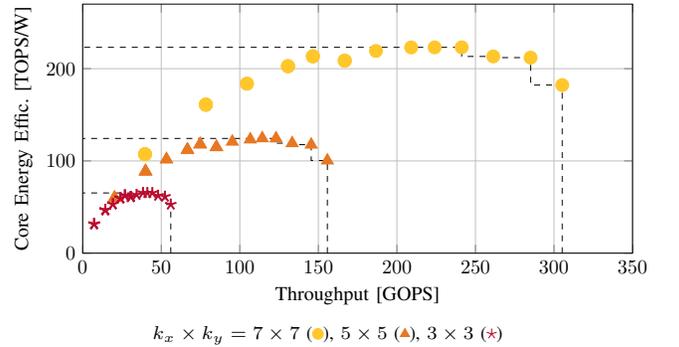\vspace{-2mm}
\section{Results \& Discussion}

\subsection{Physical Implementation}
\begin{figure}
    \centering
    \includegraphics[width=0.45\columnwidth]{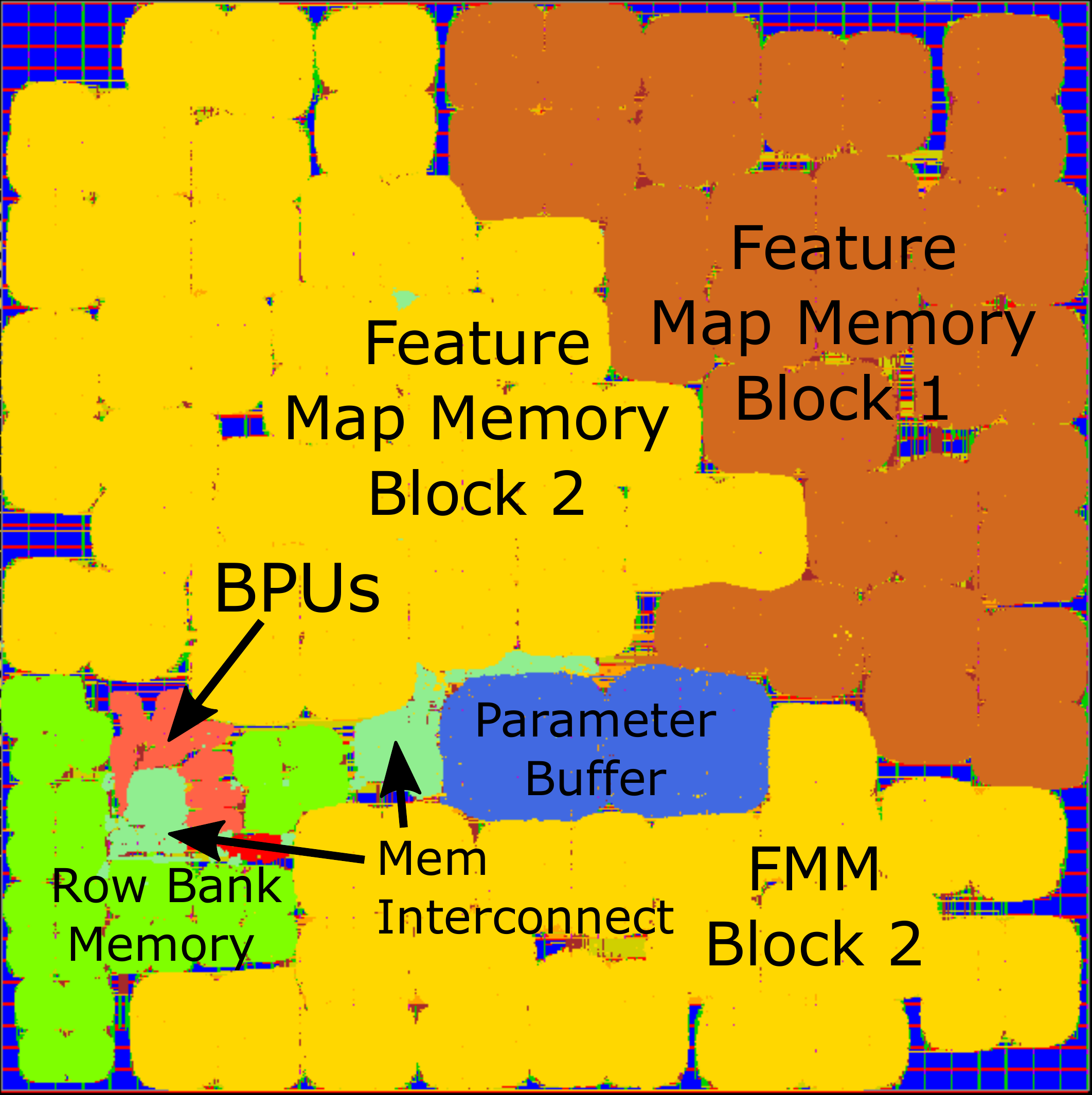}
    \includegraphics[trim={55 10 45 0}, clip, width=0.53\columnwidth]{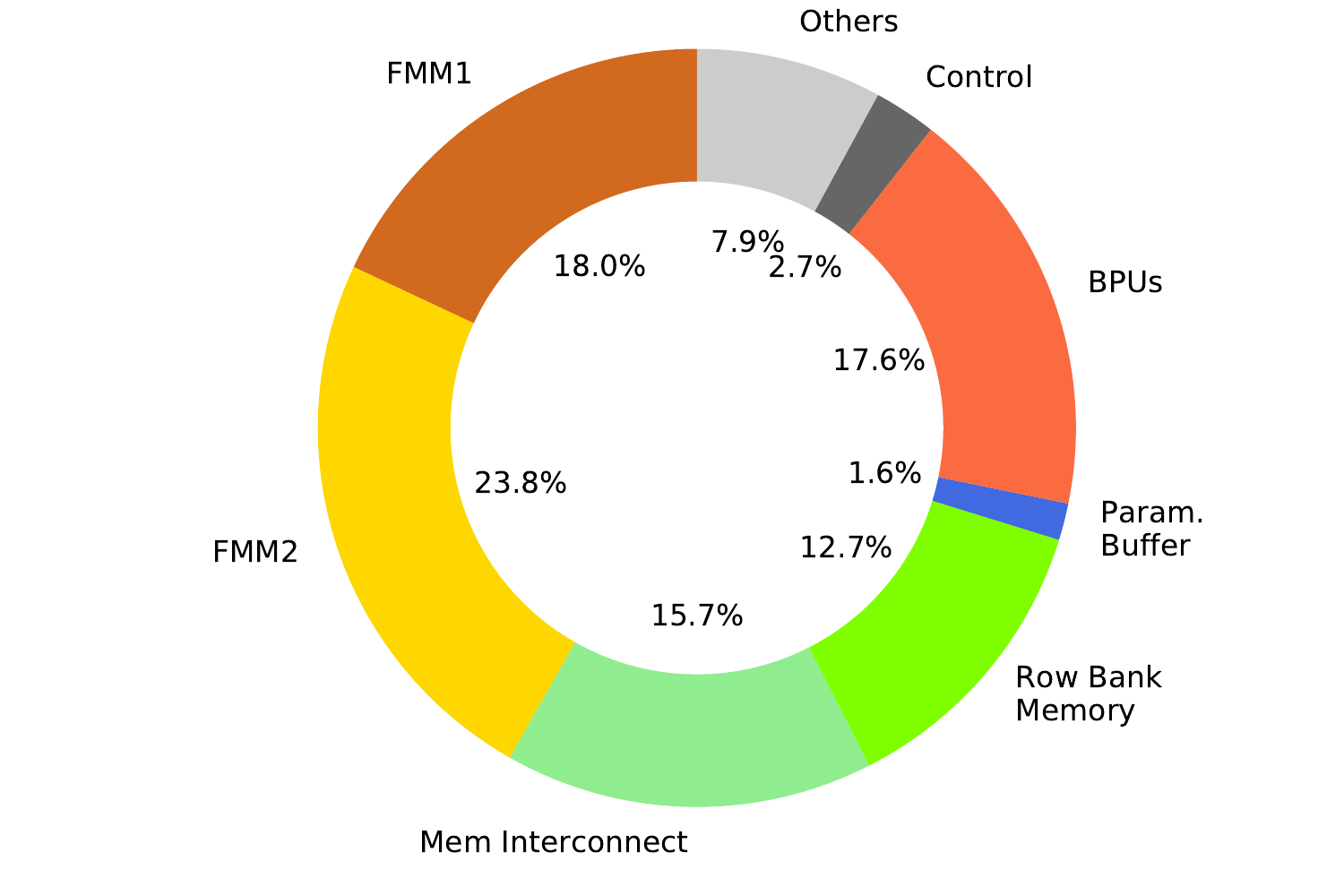}
    \caption{\iscasrev{Floorplan and power breakdown of \xnorbin core}}
    \label{fig:tsukijifloorplan}
\end{figure}

\Tsukiji{} has been implemented in GlobalFoundries 22nm FDX (7.5T). To reach the highest energy efficiency, we operate at VDD=0.4\,V with 0.1\,V forward body-biasing. To scale the voltage down to this level and to reduce the energy cost per memory access by 3.5$\times$, we use standard-cell memories (SCMs) instead of SRAMs. \addedfinal{They are designed with hierarchical clock gating and address/data silencing mechanisms, thus when a bank is not accessed the whole latch array consumes no dynamic power \cite{Andri2016a}. The SCMs are composed of multiple banks of 256 words $\times$ 32\,bit (1\,kB).} The FMM is dimensioned to fit the two largest consecutive layers\addedfinal{ of the network, which has to be supported without tiling. We have selected them to be either 16 and 32 SCM banks (48\,kB) for AlexNet or 2$\times$73 banks (146\,kB) for both AlexNet and ResNet-18}; the parameter buffer has 2 banks \addedfinal{(3.5\,kB)} and the 7 row bank memories consist of 1 SCM bank each \addedfinal{(i.e., 3.5\,kB in total)}. The final floorplan is shown in Fig.~\ref{fig:tsukijifloorplan}. It can be seen that a large part of the chip (i.e., 97\%) are memories, whereas the compute units just occupy 1\% of the total chip area of 0.7\,mm$^2$. The power consumption has been evaluated with back-annotated post-layout simulation with stimuli generated from a pretrained Torch model. \addedfinal{I/O energy has been estimated with 21\,pJ/bit (LPDDR3 memory access cost \cite{AndriHyperdriveJETCAS}).}

\subsection{Throughput to Energy Efficiency Trade-Off}
We have synthesized and run back-ends at various timing constraints, to explore the energy-efficiency to throughput trade-off at 0.4\,V, shown in Fig.~\ref{fig:xnorbinenergyvsthroughput}. Due to the lower density of the SCM memories compared to SRAMs, the chip reports high leakage, which limits the core energy efficiency to 185/100/39.0\,TOPS/W (for the full-utilization case of $7^2$/$5^2$/$3^2$ kernel sizes) with 48\,kB FMM (i.e., same size as XNORBIN) at a throughput of 241/123/44.3\,GOPS and a core power consumption of 1.3/1.2/0.90\,mW, where the power consumption can be broken down \iscasrev{as shown in Fig.~\ref{fig:tsukijifloorplan} in }\calc{1}{18.0+23.8+12.7+1.6}\%\iscasrev{\ memory, 15.7\% memory interconnects and buffers, 17.6\% compute units, 2.7\% control, and 7.9\% others.} Fortunately, most of the FMM banks stay unused and can, therefore, be power-gated. Thus, with 4 active banks (4\,kB) the energy consumption reduces to 1.08/0.99/0.68\,mW and the energy efficiency can be increased up to 223/124/65\,TOPS/W. Tab.~\ref{tab:soaacc} provides an overview of SoA BNN accelerators: the strongest analog \cite{bankman2018alwaysCORRECT}, analog-in-memory \cite{valavi201964}, and digital accelerators \cite{lee2018unpu,moons2018binareye,knag2020617}. Among the digital accelerators, UNPU is the only one supporting commonly used 3$\times$3 and 5$\times$5 kernels, while the others rely on 2$\times$2 kernels. \newiscas{Therefore, \xnorbin shows a \calc{1}{100/50.6}$\times$ better energy-efficient compared to flexible SoA BNN accelerators \cite{lee2018unpu}.} For 7$\times$7 kernels \xnorbin outperforms UNPU even \calc{1}{223/50.6}$\times$. Even though the other accelerators show a very high energy efficiency, up to 866\,TOPS/W compared to 36\,TOPS/W sustained energy efficiency for Rusci et al. \cite{rusci2018always}, \xnorbin still outperforms the SoA by \calc{1}{3.6/1.3}$\times$ in energy per classification for CIFAR-10 while achieving similar accuracies. The main reason is that they rely on uncommon $2\times 2$ kernels and have to compensate for process and ambient variations in the analog accelerators, and therefore require computationally significantly more complex networks (i.e., up to \calc{0}{2340/46.2}$\times$).
A significant improvement in accuracy can be achieved on \xnorbin by using IR-Net from Qin et al. \cite{qin2020forward}, which achieves 91.5\% accuracy at 7.3\,\textmu J per inference.

\setlength{\tabcolsep}{1mm}
\newcolumntype{H}{>{\setbox0=\hbox\bgroup}c<{\egroup}@{}}

\begin{table}[h]
{\footnotesize\caption{ \newiscas{Comparison with SoA Accelerators}. \newiscas{\xnorbin performing CIFAR-10 with: Rusci et al.\cite{rusci2018always}/IR-Net \cite{qin2020forward}}}\label{tab:soaacc}}

\begin{tabular}{@{}lrrrrrr@{}}
\toprule 
Accelerator         & \cite{valavi201964} & \cite{bankman2018alwaysCORRECT} & \cite{lee2018unpu} & \cite{moons2018binareye} & \cite{knag2020617} & OURS                 \\
Digital/Analog   & Analog/PIM           & Analog                   & Digital     & Digital                  & Digital            & Digital              \\ \midrule
Tech. Node [nm]             & 65                  & 28                       & 65          & 28                       & 10                 & 22                   \\
Area [mm\textsuperscript{2}]             & 12.6                & 4.6                      & 16          & 1.4                      & 0.39               & 0.7                  \\

MemCap [kB]          & 295                 & 328                      & 256         & 328                      & 161                & 153                  \\
Voltage [V]          & 0.68                & 0.60                     & 0.63        & 0.66                     & 0.37               & 0.40                 \\
Power [mW]           & 22                  & 0.094                    & 3.2         & 1.5                      & 5.6                & 1.1                  \\
\iscasrev{Peak} Throughput [TOPS] \hspace{-6mm}      & \textbf{19}                  & 0.072                    & 0.18       & 0.35                     & 3.4                & 0.24                 \\
\iscasrev{Peak} Efficiency [TOPS/W]\hspace{-8mm} & \textbf{866}                 & 772                      & 51          & 230                      & 617                & 223                  \\
Kernel Support   & $3^2$          & $2^2$               & $3^2,5^2$  & $2^2$               & $2^2$         & $\mathbf{1^2,3^2,5^2,7^2}$ \\
\midrule
\iscasrev{Network} &\cite{valavi201964}&\cite{moons2018binareye} &-&\cite{moons2018binareye} &\cite{moons2018binareye} &\cite{rusci2018always} / \cite{qin2020forward}\\
Accuracy CIFAR-10 [\%]\hspace{-5mm}   & 83.27               & 86.05                    & -           & 86.05                    & 86.05              & \textbf{86.8} / \textbf{91.5}               \\
Net Size [MOp] & 2'340 & 2'016 & -  & 2'016 & 2'016 &	\textbf{\calc{0}{46.2}} /\hspace{0.5mm}  \textbf{\calc{0}{175.2}}  \\
Energy/Inference [\textmu J]\hspace{-2mm}        & 3.6                 & 3.8                      & -           & 14.4                     & 5.2                & \textbf{1.3} /\ \hspace{0.3mm} 7.3                                              \\
 \bottomrule
\end{tabular}
\end{table}
\subsection{Accuracy and Energy-Efficiency for Various BNNs}
\newiscas{In this section, we compare the state-of-the-art of BNNs and how efficiently they map to \xnorbin in Tab.~\ref{tab:expres}. The networks are pretrained and there is no accuracy drop when mapping to \xnorbin due to its bit-true mapping of BNNs.
The first two BNNs have been used in embedded applications. \cerutti\ presented a BNN for Sound Event Detection on the Freesound database with 28 classes \cite{Cerutti20}. The audio data is converted to a Mel-frequency cepstral spectrogram and fed to a binary CNN with 5 layers with 3$\times$3 kernels, followed by 3 layers with 1$\times$1 kernels. They achieve a 77.9\%  accuracy (i.e., a drop of 7.9\% with respect to full-precision). Rusci et al. trained and implemented a VGG-like network on the CIFAR-10 dataset which does image recognition on 32$\times$32 colored images with 10 classes, and achieves an accuracy of 86.6\% (4.6\% less than FP32 baseline) \cite{rusci2018always}.  Both networks have been implemented on the low-power Gapuino board featuring the GAP8 multi-core processor, with 8+1 energy-optimized RISC-V cores implementing the RISC-V RV32IMC ISA and the Xpulp ISA extensions (i.e., \textit{popcount},  HW loop, post-increment lw/sw) \cite{Gautschi2017}. \addedfinal{In Cerutti et al.'s network, the input FM is tiled in 2 tiles with an overlap of 20 columns to fit in the 146\,kB FMM.} Running these networks on \xnorbin has an actual energy consumption of 1.3/353\,\textmu J/frame, a \calc{0}{760/1.3}/\calc{0}{(28.180-2.64)*1000/353}$\times$ improvement over the embedded GAP8 implementation (760\,mJ and 25.5\,J).} \newiscas{The highest BNN performance on CIFAR-10 is achieved with Group-Net based on ResNet-18, which reaches 91.5\% of accuracy (i.e., 1.5\% drop to FP32 baseline) and requires 7.3\,\textmu J.}

\newiscas{Furthermore, we evaluate the SoA BNN networks on the challenging ImageNet image classification challenge, whereas this is the first accelerator publishing performance numbers on ImageNet. DoReFa-Net with the smallest reported Top-1 accuracy gap of 12.3\% \cite{zhou2016dorefa}, XNOR-Net++ with the best Top-1 accuracy with standard BNNs \cite{bulat2019xnor}, and the two multi-base binary networks ABC-Net \cite{lin2017towards} and Zhuang et al. \cite{zhuang2019structured}. \addedfinal{DoReFa-Net was evaluated with 48\,kB FMM, and the others with 76\,kB FMM.} ABC-Net extends ResNet-18 with 3 parallel BNN layers per original full-precision layer (i.e., 3 weight bases) and reaches an accuracy of 61.0 (-8.3\%) requiring \addedfinal{1.12}\,mJ and Zhuang et al. 67.5 (\mbox{-1.8\%}) with 8$\times$ bases with a 3.0\,mJ energy cost per frame.\deletedfinal{ on the significantly larger ResNet-50} XNOR-Net++ can be run at an energy cost of 373\,\textmu J at 61\% Top-1 accuracy\deletedfinal{, and Zhuang et al. with \addedfinal{3.0}\,mJ and 67.5\% accuracy} and a throughput of 23 GOPS.}

\begin{table}[]
\scriptsize\caption{
 Real network performance on SoA BNNs. Gap shows the performance difference to the full-precision baseline network}\label{tab:expres} 

\resizebox{\columnwidth}{!}{\begin{tabular}{@{}lrrHrrrrrrr@{}}

\toprule

 & Acc. & Gap & Supported & Util.\% &  \hspace{-1mm}Core Eff.& \hspace{-1mm}Dev. Eff. & P & En. & \multicolumn{2}{r@{}}{\hspace{-1mm}Throughput}\\
& \% & \textDelta\% & Op/Op\%  & \% & \hspace{-1mm}TOPS/W & \hspace{-1mm}TOPS/W & mW & \textmu J &GOPS & FPS  \\

\multicolumn{11}{@{}l}{\textbf{Freesound}, Sound Event Detection with 28 classes, 1\x{}400\x{}64 MFCC Spectrograms}   \\

\cite{Cerutti20} 5C$^{3\times3}-$3C$^{1\times 1}$ & 77.9 & -7.9  & 48.0 & 37.9 & 4.2  & 6.5  & 2.6 & 353   & 16.8 & 7.2\\ 

\multicolumn{10}{@{}l}{\textbf{CIFAR-10}, 10 classes, 3\x{}32\x{}32} \\
\cite{rusci2018always} VGG-like         & 86.8 & -4.6  & x    & 60.9 & 36.1 & 8.2  & 2.9 & 1.3   & 23.9 & 2286.2 \\  
\cite{qin2020forward} ResNet-18 & 91.5 & -1.5 & 92.6 & 49.2 & 24.0 & 9.2  & 2.4 & 7.3   & 21.8 & 324.3 \\

\multicolumn{8}{@{}l}{\textbf{ILSVRC}, 1'000 classes, 3\x{}224\x{}224} \\
\cite{zhou2016dorefa} AlexNet                   & 43.6 & -12.3 & 85.3 & 45.3 & 27.8 & 13.8 & 2.1 & 141   & 28.5 & 14.7                 \\
\cite{bulat2019xnor} ResNet-18                  & 57.1 & -12.2 & 93.4 & 52.9 & 24.7 & 9.0  & 2.6 & 373   & 23.4 & 7.0           \\
\cite{lin2017towards} 3$\times$ ResNet-18       & 61.0 & -8.3  & 93.4 & 52.9 & 24.7 & 9.0  & 2.6 & 1,118 & 23.4 & 2.3                 \\
\cite{zhuang2019structured} 8$\times$ ResNet-18 & 67.5 & -1.8  & 93.4 & 52.9 & 24.7 & 9.0  & 2.6 & 2,981 & 23.4 & 0.9                        \\

\bottomrule
\end{tabular}}
\end{table}

\section{Conclusion}
We have presented a novel BNN accelerator implemented in GlobalFoundries 22\,nm technology, which can perform inference with 2.8$\times{}$ less energy on CIFAR-10 at the same accuracy than the current state-of-the-art, an analog processing-in-memory device, and 4$\times{}$ less energy than the next best digital solution which relies on the more recent 10\,nm node---all the while  providing the option of running also higher accuracy BNNs out of reach of current analog computing architectures. We achieve this by combining a high energy efficiency of up to 223\,TOPS/W, attributable to an efficient on-chip memory hierarchy based on SCMs and strong voltage scaling, with the flexibility to run various filter kernel sizes that allows us to use  more efficient BNNs without sacrificing accuracy, thereby boosting the overall energy efficiency beyond what can be measured by the simple TOPS/W metric. 
Further, we provide the first performance numbers for BNN hardware accelerators on the ILSVRC dataset, as we can run state-of-the-art BNNs such as XNOR-Net++ or Group-Net, reaching a Top-1 accuracy of 61.0\% and 67.5\% on the ILSVRC'12 dataset at merely 1.1 and 3.0\,mJ/frame, respectively.

\bibliographystyle{IEEEtran}
\bibliography{myBstCtl, refs, refspecial}

\end{document}